**Multi-replica biased sampling for photoswitchable $\pi$-conjugated polymers**

Mariagrazia Fortino[1], Concetta Cozza[1], Massimiliano Bonomi[2,*], Adriana Pietropaolo[3,*]

**AFFILIATIONS**

[1] Dipartimento di Scienze della Salute, Università di Catanzaro, Viale Europa, 88100 Catanzaro, Italy.
[2] Structural Bioinformatics Unit, Department of Structural Biology and Chemistry; CNRS UMR 3528; C3BI, CNRS USR 3756; Institut Pasteur, Paris, France. ORCID: 0000-0002-7321-0004.
[3] Dipartimento di Scienze della Salute, Università di Catanzaro, Viale Europa, 88100 Catanzaro, Italy. ORCID: 0000-0003-0955-2058

[*]Authors to whom correspondence should be addressed:
apietropaolo@unicz.it, mbonomi@pasteur.fr

**ABSTRACT**

In recent years, $\pi$-conjugated polymers are attracting considerable interest in view of their light-dependent torsional reorganization around the $\pi$-conjugated backbone, which determines peculiar light-emitting properties. Motivated by the interest in designing conjugated polymers with tunable photoswitchable pathways, we devised a computational framework to enhance the sampling of the torsional conformational space and at the same time estimate ground to excited-state free-energy differences. This scheme is based on a combination of Hamiltonian Replica Exchange (REM), Parallel Bias metadynamics, and free-energy perturbation theory. In our scheme, each REM replica samples an intermediate unphysical state between the ground and the first two excited states, which are characterized by TD-DFT simulations at the B3LYP/6-31G* level of theory. We applied the method to a 5-mer of 9,9-dioctylfluorene and found that upon irradiation this system can undergo a dihedral inversion from -155 to 155 degrees crossing a barrier that decreases from 0.1 eV in the ground state ($S_0$) to 0.05 eV and 0.04 eV in the first ($S_1$) and second ($S_2$) excited states. Furthermore, $S_1$ and even more $S_2$ were predicted to stabilize coplanar dihedrals, with a local free-energy minimum located at $\pm 44$ degrees. The presence of a free-energy barrier of 0.08 eV for the $S_1$ and 0.12 eV for the $S_2$ state can trap this conformation in a basin far from the global free-energy minimum located at 155 degrees. The simulation results were compared with the experimental emission spectrum, showing a quantitative agreement with the predictions provided by our framework.

**I. INTRODUCTION**

Conjugated polymers are a class of organic frameworks that are widely used in an extensive range of electronic applications, thanks to their promising optical and electronic properties.[1] These polymers offer an excellent alternative to inorganic materials since they are moderately low-cost.[2] For these reasons, conjugated polymers have found remarkable applications in devices such as organic photovoltaics (OPV), organic light-emitting diodes (OLED), organic field-effect transistors (OFET) and a variety of sensors.[3-12] Polyfluorene derivatives are of wide interest in the field of organic electronics. A peculiarity of these systems is the photo-induced torsional reorganization around the conjugated backbone,[13,14] through a *twisted-coplanar* transition, which allows a specific molecular response after light-irradiation.[15-18] Several computational efforts have been performed in the last decades to predict the properties of polyfluorenes and ultimately to optimize their structural[19] and optical properties,[20,21] electronic structure[22-24] or charge-transport character.[25]

In 2012, *Clark et al.* have shown through non-adiabatic excited-state dynamic simulations that the photo-induced torsional relaxation around the backbone of 5-mer of 9,9-dioctylfluorene may occur in a toluene solution in a sub-100 fs timescale.[26] A flattening mechanism was predicted to occur during the decay, showing the conversion of the electronic potential energy into torsional kinetic energy. However, non-adiabatic excited state molecular dynamics simulations often require excited-state quantum simulations performed at every time step on an ensemble of trajectories, each propagated for a picosecond or more. Consequently, although this field has matured significantly with remarkable improvements in the simulation of very large molecular systems, it often requires to make compromises in order to find the best balance between precision and numerical cost.[27] Force-field based excited state simulations are promising for studying large-scale transitions occurring during photo irradiation.[28-30] Various simulation frameworks have been proposed in order to derive classical force fields able to accurately describe molecules in the ground and excited states.[31-33] This is a field of active research and, for example, inclusion of more complex functional forms including the torsion dihedrals might be important for accurate studies of photo-induced processes.

Recently, we have developed a simple approach[34] to estimate the free-energy gap between ground and excited states of a fluorene pentamer, while at the same time enhancing the sampling of different conformers. Within this simulation framework, several sets of independent free-energy simulations are performed to enable an exhaustive sampling of the conformational space, and free-energy perturbation theory (FEP)[35,36] is then used to provide an accurate estimate of the free-energy gap between ground and first excited state. In this work, we extend the method by biasing multiple torsions of fluorene-based oligomers substituted in the 9-position by dioctyl chains (Scheme 1), a substitution that is usually employed to improve the solubility in organic solvents. The Parallel Bias metadynamics approach[37] combined with Hamiltonian Replica Exchange (REM)[38] are used to increase the efficiency of our scheme and, in combination with FEP, characterize the free-energy landscapes of the first singlet $S_0$ ground state and the lowest singlet excited states $S_1$ and $S_2$ as well as the transitions between these states.

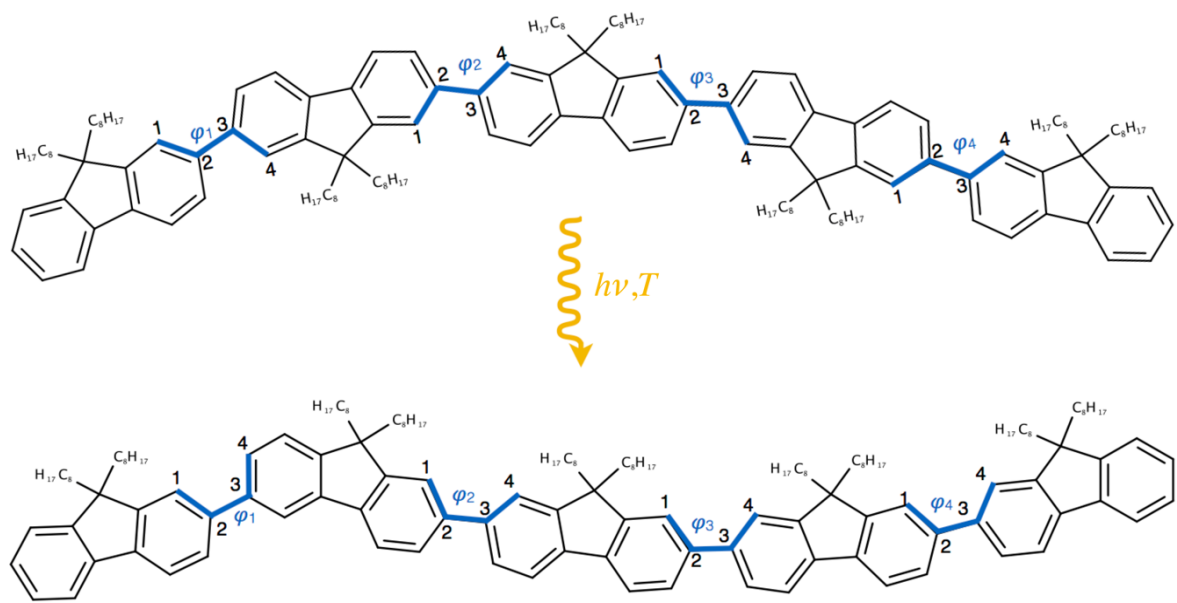

**Scheme 1.** Photo-induced twisted-coplanar transition occurring on a 5-mer of 9,9-dioctylfluorene.

## II. THEORETICAL FRAMEWORK

*Derivation of the torsional potential*

The potential energy surfaces (PES) as a function of the dihedral rotation were performed within the Density Functional Theory (DFT) framework for the torsions belonging of a fluorene pentamer in the ground and in the first two singlet excited states. The Becke's three-parameter hybrid functional (B3) with the Lee, Yang, and Parr (LYP) expression for the nonlocal correlation, B3LYP[39] were initially used and compared with the Coulomb-attenuating method CAM-B3LYP.[40] We compared the PES at the CAM-B3LYP/6-31G* and B3LYP/6-31G* levels of theory. We detected a shift of 0.62 eV in the $S_1$ state higher for the CAM-B3LYP/6-31G* and a rising of the coplanar barrier of 0.02 eV and 0.05 eV, respectively in $S_0$ and $S_1$ states, with respect to the PES calculated at the B3LYP/6-31G* level of theory. Therefore, we proceeded to compute the PES at the B3LYP/6-31G* level of theory using Gaussian16,[41] in line with previous reports indicating this level of theory is suitable to predict the electronic structure of polyfluorenes.[22]

The torsional conformers of the fluorene pentamer were optimized by varying each of the $\varphi_1, \varphi_2, \varphi_3$ torsions from 0 to 180 degrees in the $S_0$, $S_1$ and $S_2$ states. Due to the $C_2$ symmetric repeating unit, $\varphi_1$ is equivalent to $\varphi_4$ and $\varphi_2$ to $\varphi_3$. While in the ground state the $E^{QM}$ values of the PES of the consecutive torsions does not show appreciable differences, in the excited states the external ($\varphi_1$) and the internal torsion ($\varphi_2$) or ($\varphi_3$) markedly vary (Figure 1 and Figure S6-S8). In particular, the PES of the second torsion ($\varphi_2$) shows lower energy values for coplanar conformers and a higher potential energy barrier at 90 degrees that increases owing to the loss of conjugation, with respect to the PES of the first torsion ($\varphi_1$). Coplanar states are stabilized through light irradiation owing to the lowering of the potential energy barriers at 0 degrees in $S_1$ and $S_2$ excited states and to the increase of the potential energy barrier at 90 degrees. Consequently, we derived two sets of torsional potentials for the external and the internal torsions of the polymer. Non-linear least-square interpolations were performed by minimizing the square difference $\Delta E$ between the molecular mechanics energies ($E^{MM}$) and the *ab initio* energies ($E^{QM}$) for the $N$ points on the PES of the electronic state *i*, reported in Eq. 1.

$$\Delta E_i = \sum_{j=1}^{N}(E^{QMi}(\varphi_j) - E^{MMi}(\varphi_j))^2 \tag{1}$$

where:

$$E^{MMi}(\varphi) = E^{CHARMM} + V^{S_i}(\varphi) + \sum K_m(1 + \cos(m\phi - \phi_0)) \tag{2}$$

and:

$$V^{S_i}(\varphi) = \sum_{n=0}^{5} C_n^{S_i}(\cos(\varphi - 180))^n \tag{3}$$

in which $E^{CHARMM}$ represents the force-field potential CHARMM CGenFF[42] and $V^{S_i}(\varphi)$ represents the torsional potential in a given electronic state *i* related to the torsions for which each PES is calculated ($\varphi$ is defined as *1234* from Scheme 1).

The term $K_m$ and $\phi_0$ represent the force constants and the phase of a cosine series describing the other proper $\phi$ dihedral acting on the fluorene-fluorene rotation. The values of $E^{MM}$ were calculated from an energy minimization of the optimized coordinates of the QM PES. We also tested the fitting parameters through the torsional free-energy profiles calculated at 200 K, where the free energy approaches the PES (Figure 1). The $K_m$ and $\phi_0$ terms of the proper dihedrals were calculated with iterative fits starting from $K_m$ =0 and using as initial guess the fitting parameters from the QM PES interpolation for the $\varphi$ torsion in a given electronic state. As shown in the free-energy profiles reconstructed at 200 K, together with the potential energy minimization (the oscillations in the angles mostly derived from the minimization algorithm), the $K_m$ and $\phi_0$ terms found for the ground-state combined with the fitting parameters of the QM PES in a given electronic state, well reproduced the potential energy along the $\varphi$ torsion (Figure 1).

The fitting parameters describing the $V^{S_i}(\varphi)$ potential are reported in Table 1, the parameters describing the proper dihedrals are reported in Table 2. The ESP charges were calculated for each optimized electronic state of the 5-mer oligofluorene at the B3LYP/6-31G* and are reported in Tables S1 of Supplementary Material. We also calculated the ESP charges at the MP2[43] and CIS[44] level of theory. Those are consistent with the ones derived at the B3LYP/6-31G* level of theory and are reported in Table S2 of the Supplementary Material.

**Table 1**. The coefficients $C_n^{S_i}$ (kJ/mol) of the Ryckaert-Bellemans function described in Eq. 3 and used in GROMACS, obtained through a non-linear least-square interpolation of the PES values obtained at the B3LYP/6-31G* level.

| C0 | C1 | C2 | C3 | C4 | C5 |
|---|---|---|---|---|---|
| **$S_0(\varphi)$** | | | | | |
| 12.16 | -0.02400 | -43.32 | 0.1200 | 38.80 | -0.06010 |
| **$S_1(\varphi_1)$** | | | | | |
| 301.3 | -0.1478 | -48.16 | 0.9948 | 34.30 | -1.001 |
| **$S_2(\varphi_1)$** | | | | | |
| 348.8 | -0.1104 | -54.71 | 0.7533 | 35.14 | -0.8068 |
| **$S_1(\varphi_2)$** | | | | | |
| 313.9 | 0.3972 | -72.12 | -0.9463 | 43.87 | 0.5858 |
| **$S_2(\varphi_2)$** | | | | | |
| 351.9 | -0.7135 | -69.01 | 0.653 | 47.11 | 0.41 |

**Table 2.** Torsional parameters for the proper dihedrals $\phi$ described in Eq. 2, including one multiple cosine function and two single cosine functions.

| $K_m$ (kJ/mol) | $\phi_0$ | m |
|---|---|---|
| -6.40 | 180 | 4 |
| 2.20 | 180 | 2 |
| -8.20 | 0.0 | 4 |
| | | |
| -2.20 | 0 | 2 |
| | | |
| -2.20 | 0 | 2 |

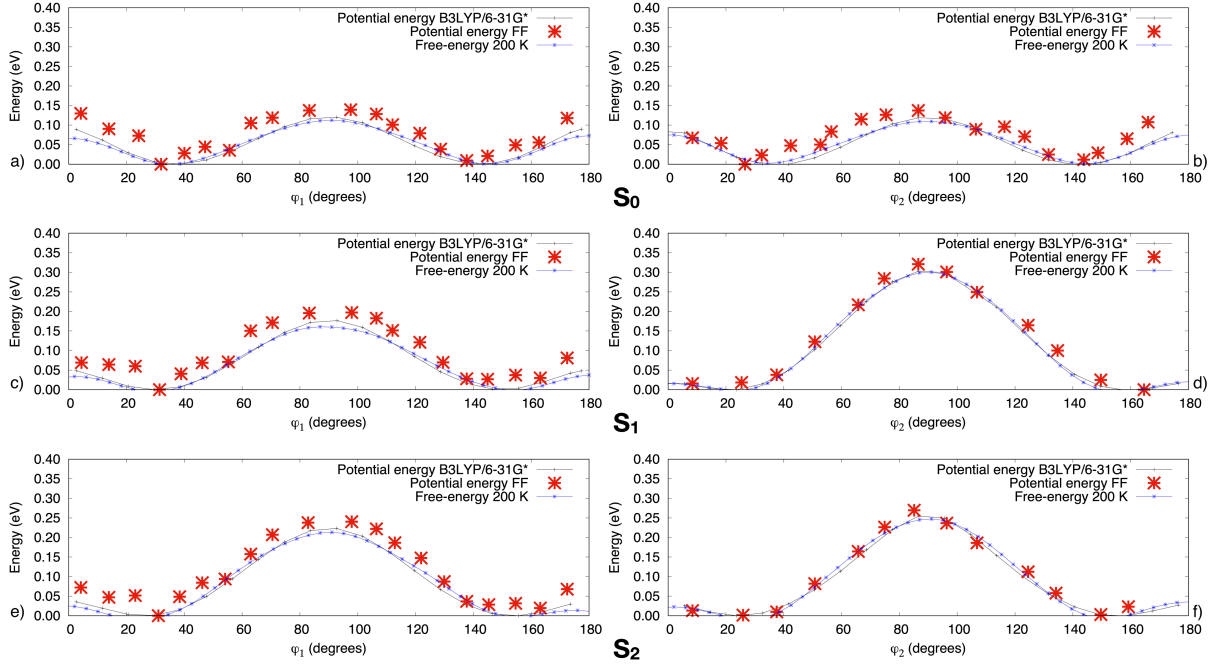

**Figure 1.** Fluorene-fluorene torsional potential calculated from energy-minimized conformations at MM level and from free energy calculated with Well-tempered metadynamics simulations at 200 K. The total energy calculated at B3LYP/6-31G* level of theory is also shown. a) Calculations related to the torsion ($\varphi_1$) of the fluorene pentamer in the $S_0$ state, b) calculations related to the torsion ($\varphi_2$) of the fluorene pentamer in the $S_0$ state, c) calculations related to the torsion ($\varphi_1$) of the fluorene pentamer in the $S_1$ state, d) calculations related to the torsion ($\varphi_2$) of the fluorene pentamer in the $S_1$ state, e) calculations related to the torsion ($\varphi_1$) of the fluorene pentamer in the $S_2$ state, f) calculations related to the torsion ($\varphi_2$) of the fluorene pentamer in the $S_2$ state.

*Hamiltonian Replica Exchange combined with Free-energy perturbation*
The free-energy as a function of the fluorene-fluorene dihedral angles $\varphi_i$ is defined as follows:

$$F^{S_i}(\varphi_i) = -\frac{1}{\beta}\log\left(\frac{\int \delta(\varphi_i - \varphi_i(r)) e^{-\beta V(r)} dr}{\int e^{-\beta V(r)} dr}\right) \quad (4)$$

In this equation, $V(r)$ represents the interatomic potential (the force field), $r$ the atomic coordinates, and $\beta$ represents the inverse of $k_B T$, where $k_B$ is the Boltzmann constant and $T$ the temperature of the system. We now add to the interatomic potential the torsional potential reported $V^{S_i}(\varphi_i)$ defined in Eq.3. Hence, the free-energy surface for a given electronic state $F^{S_i}(\varphi_i)$ becomes:

$$F^{S_i}(\varphi_i) = -\frac{1}{\beta}\log\left(\frac{\int \delta(\varphi_i - \varphi_i(r)) e^{-\beta\left(V(r) + V^{S_i}(\varphi_i(r))\right)} dr}{\int e^{-\beta\left(V(r) + V^{S_i}(\varphi_i(r))\right)} dr}\right) \quad (5)$$

In order to evaluate the free-energy difference between the $S_0$, $S_1$ and $S_2$ electronic states, FEP was used in combination with REM, following the approach introduced in Ref.[45]. Multiple intermediate windows between the $S_0$, $S_1$ and $S_2$ electronic states were introduced through two parameters $\lambda_1$ and $\lambda_2$. The former is used to reach $S_1$ from $S_0$ ($\lambda_1: 0 \rightarrow 1; \lambda_2 = 0$); the latter to reach $S_2$ from $S_1$ ($\lambda_1 = 1; \lambda_2: 0 \rightarrow 1$) (Scheme 2).

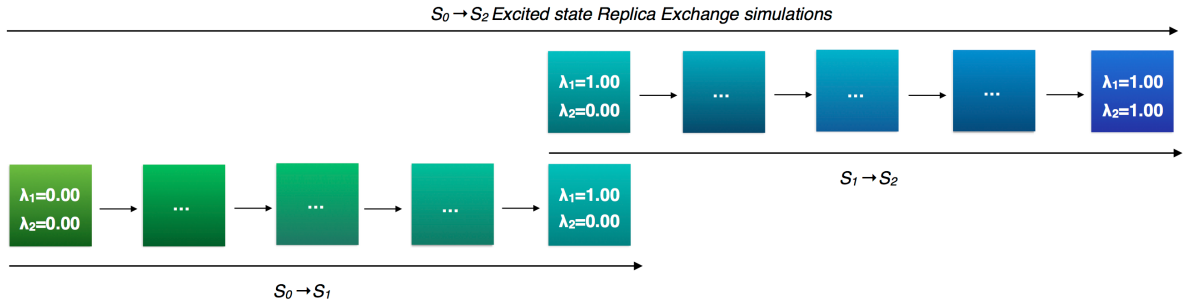

**Scheme 2.** Scheme of the simulation framework herein proposed based on Hamiltonian Replica Exchange, where each replica differs from the torsional potentials through $\lambda_1$ and $\lambda_2$ parameters. Each torsion is biased within the parallel bias metadynamics approach.

The external potential acting on each replica is defined as:

$$V_\lambda(\varphi) = \sum_{m=1}^{4} \left[ (1-\lambda_1)V^{S_0}(\varphi_m) + \lambda_1(1-\lambda_2)V^{S_1}(\varphi_m) + \lambda_2 V^{S_2}(\varphi_m) \right] \qquad 6)$$

Where $m$ represents the number of torsions belonging to the polymer. In order to obtain an adequate overlap between the energy distributions of adjacent windows, a step sizes $\Delta\lambda_1$ equal to 0.04 and $\Delta\lambda_2$ equal to 0.1 were used for the $S_0$-$S_1$ and $S_1$-$S_2$ transitions, respectively. The step size was chosen as a compromise between achieving an overlap between torsional potentials of neighboring replicas and keeping the simulation cost moderate (i.e. the minimum number of $\lambda$ windows to allow an overlap between torsional potentials).

In this REM framework, $N$ non-interacting replicas of the system, each identified by a given pair $\lambda = (\lambda_1,\lambda_2)$ of control parameters, were simulated in parallel. At fixed time intervals, a swap between two configurations belonging to neighboring replicas was attempted and accepted using a Metropolis criterion based on the overlap between energy distributions of the two replicas. The use of the different $(\lambda_1,\lambda_2)$ intermediate windows makes possible an efficient exchange between replicas. In this way, instead of multiple independent FEP runs, the simulation can be executed in a single multi-replica run. Within this framework, the ground state replica with higher coplanar free-energy barriers can be exchanged with excited state replica having lower coplanar free-energy barriers, thus enhancing sampling. Furthermore, the overlap between energy distributions of neighboring replicas, which is ensured to obtain efficient diffusion in the $\lambda = (\lambda_1,\lambda_2)$ space, guarantees also an accurate calculation of the free-energy difference between two neighboring replicas using the FEP approach in Eq. 7:

$$\Delta F^{(\lambda_i+\Delta\lambda)} = -\frac{1}{\beta}\log\langle e^{-\beta(V_{\lambda_i+\Delta\lambda}-V_{\lambda_i})}\rangle_{\lambda_i} \tag{7}$$

Finally, the free-energy gap between the $S_0$ and $S_1$ and $S_1$ and $S_2$ electronic states can be estimated as the sum over the free-energy differences between neighboring replicas using Eq. 8:

$$\Delta F^{S_i \to S_{i+1}} = \sum_i \Delta F^{(\lambda_i+\Delta\lambda)} \tag{8}$$

where the free-energy gap has been divided by the number of torsions.

*Enhanced sampling with Parallel Bias metadynamics.*

Parallel Bias metadynamics (PBMetaD)[37] was used to enhance sampling in each REM replica. The four main-chain dihedrals belonging to the polymer ($\varphi_1, \varphi_2, \varphi_3, \varphi_4$) were used as Collective Variables (CVs). Multiple mono-dimensional metadynamics[46] bias potentials $V_G(\varphi_i; t)$ were used to construct the PBMetaD bias potential $V_{PB}(\varphi_1, \varphi_2, \varphi_3, \varphi_4; t)$ defined by:

$$V_{PB}(\varphi_1, \varphi_2, \varphi_3, \varphi_4; t) = -\frac{1}{\beta}\log\left(\sum_{i=1}^{4} e^{-\beta V_G(\varphi_i;t)}\right) \tag{9}$$

The individual metadynamics potential $V_G(\varphi_i; t)$ is built adaptively during the course of the simulation by depositing Gaussian functions along the system trajectory in the CVs space:

$$V_G(\varphi_i; t) = \int_0^t dt' \, \omega(t') \cdot \exp\left(-\frac{\left(\varphi_i(r)-\varphi_i(r(t'))\right)^2}{2\sigma_i^2}\right) \tag{10}$$

where $t$ is simulation time, $\sigma_i$ the Gaussian width of the *i*-th CV, and $\omega$ the deposition rate of the bias potential. Typically, Gaussians are added to the simulation with a discrete and constant deposition stride $\tau$. Therefore, the deposition rate $\omega$ is expressed as the ratio between the Gaussian height $W$ and the deposition stride $\tau$. Similarly to well-tempered metadynamics,[47] in PBMetaD the Gaussian height decreases with simulation time as:

$$W(t) = W_0 \cdot \exp\left(-\frac{V_G(\varphi_i;t)}{k_B\Delta T}\right) \cdot \frac{e^{-\beta V_G(\varphi_i;t)}}{\sum_{i=1}^{4} e^{-\beta V_G(\varphi_i;t)}} \tag{11}$$

where $W_0$ is the initial Gaussian height and $\Delta T$ an input parameter with the dimension of a temperature.

$\Delta T$ can be used to limit the exploration to relevant regions of the CV space and thus avoid visiting excessively high free-energy areas. This parameter is often expressed in terms of the so-called biasfactor:

$$\gamma = \frac{T+\Delta T}{T} \tag{12}$$

At convergence, one-dimensional free-energy profiles $F(\varphi_i)$ can be reconstructed directly from the bias potentials $V_G(\varphi_i; t)$ as in standard well-tempered metadynamics simulations:

$$V_G(\varphi_i; t \rightarrow \infty) = -\frac{\Delta T}{T+\Delta T} \cdot F(\varphi_i) + C \tag{13}$$

while multi-dimensional free-energy surfaces can be reconstructed by reweighting.[48] The presence of the PBMetaD bias potential $V_{PB}(\varphi_1, \varphi_2, \varphi_3, \varphi_4; t)$ was accounted for when calculating the swap acceptance probability in the REM scheme as well as when computing the FEP ensemble average in Eq. 7.

*Simulation Details*

The initial 5-mer of 9,9-dioctylfluorene structures were built by selecting the polymeric chains from the simulated system reported in Ref.[49]

In order to derive the force field in the ground and excited states, potential energy scans as a function of the dihedral rotation of the unsubstituted fluorene pentamer were carried out using the Density Functional Theory (DFT) framework with the B3LYP[39] functionals and the 6-31G* basis set within the Gaussian16 package.[41] The torsional conformers were optimized varying the torsion from 0 to 180 degrees in the $S_0$ and $S_1$ states.

Subsequently, the initial coordinates were energy-minimized *in vacuo* using the Steepest Descent Algorithm and equilibrated *in vacuo* at 200 K and at 300 K using the *ab initio* derived torsional potential combined with the CHARMM based General Force Field.[42] The equilibrated conformation at 200 K was used as starting coordinates for reconstructing the free-energy profile at 200 K. Specifically, two well-tempered metadynamics[47] runs were carried out using the first (external) or the second (inner) polyfluorene dihedrals. Gaussians with initial height equal to 1.2 kJ/mol and with width of 0.2 radians, were deposited with a stride of 1 ps and biasfactor equal to 25 in order to allow an efficient diffusion among replicas, with a temperature of 300 K.

The equilibrated conformation at 300 K was used as starting coordinates for the REM simulations. 36 replicas were used: 26 to cover the interval $S_0$, $S_1$ and 10 for the interval $S_1$,

$S_2$. Each replica was simulated in the NVT ensemble at 300 K enforced by the velocity rescaling thermostat.[50]

A swap between two neighboring replicas was attempted every 100 MD steps and accepted using a Metropolis criterion. For the PBMetaD approach, Gaussians with initial height equal to 1.2 kJ/mol were deposited with a stride of 1 ps and biasfactor equal to 40. The simulation length of each replica was 0.085 µs, for a total aggregated simulation time of 3.06 µs.

In all simulations, the time step was set at 2 fs. The LINCS algorithm[51] was used to fix all bonds lengths. For the Lennard-Jones and electrostatic interactions, a cutoff of 2.0 nm was used. The Particle Mesh Ewald method[52] was used to calculate electrostatic interactions. Periodic boundary conditions were applied using a cell dimension of 48.140x64.085x16.445 nm$^3$ with a system size of 347 atoms. All simulations were performed with GROMACS 2020.2[53] equipped with PLUMED 2.7.0.[54] The conformations sampled in the $S_1$ state have been clustered with torsional angles ranging from 20 to 155 degrees. The oscillator strengths have been computed for those selected coordinates using the ZIndo/S method[55,56] as implemented in Gaussian 16 package,[41] and subsequently scaled with the square of each wavenumbers. The calculations of the emission spectra at a given emission energy were carried out assuming Gaussian bands with 50 cm$^{-1}$ full width at half-height for all transitions centered in a given emission energy converted in wavenumbers.

### III. FREE-ENERGY PATHWAYS IN THE GROUND AND EXCITED STATES

The free-energy profiles reconstructed for a 5-mer of 9,9-dioctylfluorene as a function of the dihedral angles ($\varphi_1$, $\varphi_2$, $\varphi_3$, $\varphi_4$) and concerning the $S_0$, $S_1$ and $S_2$ states are shown in Figure 2. Inspection of the simulated free-energy profiles at 300 K pointed out striking differences with the adiabatic PES computed at the *ab-initio* level (shown in Figure 1). First, the minima at 44 degrees and 155 degrees, which were predicted to be degenerate in the former PES (Figure 1), showed in the free-energy profiles different stabilities of 0.14 eV in the $S_0$ state and 0.15 eV in $S_1$ and $S_2$ states (Figure 2). Second, the coplanar barrier decreased by 0.05 eV in $S_1$ and 0.06 eV in $S_2$ free-energy profiles with respect to those of the ground state (Figure 2). No significant differences have been observed in the free-energy profiles of the ground state as a function of the $\varphi_1$, $\varphi_2$, $\varphi_3$, $\varphi_4$ dihedrals.

Small variations appear in the free-energy profiles of the excited states. The $S_1$ state shows different orthogonal free-energy barrier at 90 degrees on the external and internal torsions (0.22 eV for $\varphi_1$, 0.34 eV for $\varphi_2$, 0.33 eV for $\varphi_3$ and 0.24 for $\varphi_4$), whereas the $S_2$ state shows variations in the coplanar free-energy barrier (0.03 eV for $\varphi_1$ or $\varphi_4$; 0.09 eV for $\varphi_2$ and 0.11 eV for $\varphi_3$).

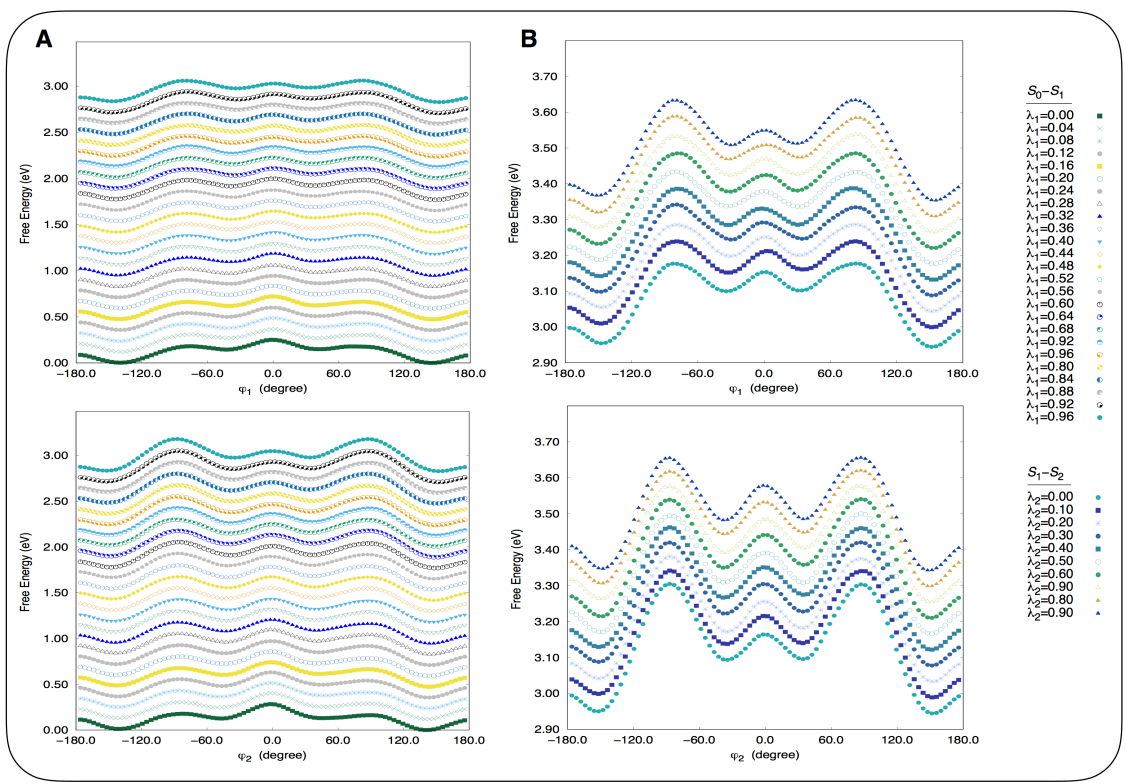

**Figure 2. A.** Free-energy profiles computed using the Parallel Bias metadynamics approach combined with Hamiltonian Replica Exchange, as a function of the dihedral angles $\varphi_1$, $\varphi_2$ and for the $S_1$-$S_0$ transition. **B.** Free-energy profiles for the $S_2$-$S_1$ transition as a function of the dihedral angles $\varphi_1, \varphi_2$.

The calculated $S_1$-$S_0$ free-energy gap estimated by FEP is equal to 2.946 eV, while the calculated $S_1$-$S_2$ free energy gap is predicted as 0.442 eV, both with a statistical error of $2\times10^{-4}$ eV. Notably, the calculated $S_1$-$S_0$ free-energy gap shows a good agreement with the experimental photoluminescence spectrum centered at 2.88 eV reported for a 5-mer of 9,9-dioctylfluorene.[57] Since we derived relaxed QM PES, we predicted emission free-energy estimates and those turn out to be slightly higher than the experimental one ($2.946 \pm 2\times10^{-4}$ eV vs 2.88 eV). Indeed, this protocol can be straightforwardly extended to absorption processes, by fitting a Franck-Condon vertical transition from the ground state instead of a relaxed PES, as in this case.

At variance with an unsubstituted 5-mer oligofluorene in which the free-energy landscape is characterized by four degenerate free-energy minima,[34] the dioctyl chains stabilize the dihedral values of 155 degrees close to the values of 144 degrees typical of a $5_2$ helix, instead of the values of 72 degrees typical of a $5_1$ helix. The more favorable packing for a $5_2$ helix instead of a $5_1$ helix was previously suggested from electron diffraction patters combined with geometry optimizations at RHF/6-31G(d,p) level of theory.[58] The $5_2$ helix is further stabilized in the $S_1$ and $S_2$ excited states where a higher orthogonal free-energy barrier at 90 degrees is observed for the free-energy profiles reconstructed along the first $\varphi_1$ dihedral of $S_1$ and $S_2$ states (0.23 eV and 0.28 eV, respectively) compared to the free-energy barrier of 0.18 eV in the one of $S_0$.

A *switch* region is detected at values of angles of 44 degrees, where a reduced free-energy coplanar barrier is observed for the $\varphi_1$ transition towards a coplanar state (~0.04 eV) with respect to that belonging to the ground state (~0.10 eV). Notably, the dihedral angles of the absolute free-energy basins in $S_1$ and $S_2$ states are shifted towards higher values approaching 155 degrees, whereas the dihedral angles of the *switch* free-energy basin are stabilized towards values approaching 27 degrees (Figure 2), favoring in this way a molecular switch from positive to coplanar dihedrals reaching negative twist dihedrals (Figure 3). These results are in agreement with the typical twisted-coplanar transition of biphenyl-based rotors, herein represented in the molecular conformations sketched in Figure 3 for $S_0$, $S_1$ and $S_2$ electronic states. In particular, the free-energy surfaces reconstructed for $S_0$, $S_1$ and $S_2$ states as a function of two consecutive dihedrals are highlighted in Figure 3, together with the relative switching pathway. From those isosurfaces, the minimal free-energy pathway is predicted to occur through the torsion periodicity at 180 degrees, allowing the dihedral inversion of the most stable conformer with 155 degrees (path from A to C). The barriers at 180 degrees progressively decrease going from the $S_0$ to $S_1$ and $S_2$ states, as highlighted in the monodimensional free-energy profiles reconstructed aligning the horizontal (C-B transition) and vertical (A-B transition) paths along $\varphi_1$ and $\varphi_2$ or $\varphi_2$ and $\varphi_3$ (Figure 3). Noteworthy, $S_1$ and even more $S_2$ were predicted to stabilize coplanar dihedrals, with a local free-energy minimum located at ±44 degrees far from the global free-energy minimum at 155 degrees.

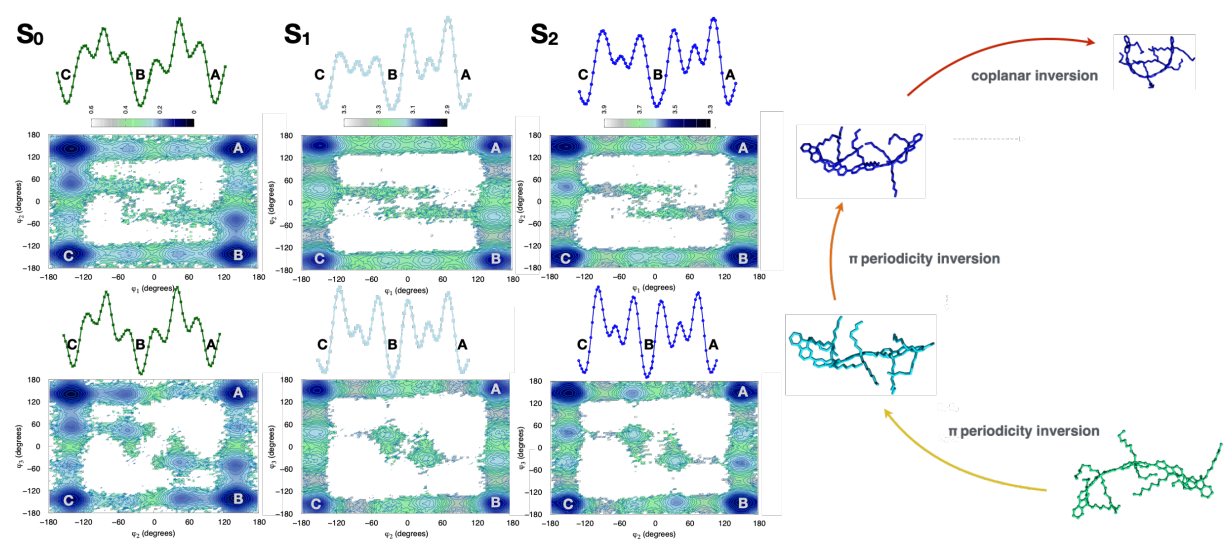

**Figure 3.** Free-energy isosurfaces reconstructed as a function of the $\varphi_1$ and $\varphi_2$ dihedrals, with the monodimensional free-energy profiles representing C-B and A-B transition. The photoswitchable mechanism from S0 to S2 is shown and predicted with a π-periodicity inversion.

We finally assessed the proposed mechanism through the calculation of the emission spectrum (Figure 4). The predicted emission spectrum, despite being shifted by 0.06 eV with respect to the experimental profile reported by Kang et al.,[57] presents a similar shape with two main central peaks. Our multidimensional free-energy surfaces suggest a photoswitchable transition starting from the free-energy minimum approaching the $5_2$ helix conformation. Upon irradiation the 5-mer of 9,9-dioctylfluorene can invert its dihedrals through the π-periodicity. A further transition can undergo away from the global free-energy minimum in the $S_1$ and in $S_2$ state, where the coplanar dihedrals and the conformers located at ±44 degrees are stabilized by higher free-energy barriers (0.03 eV for $S_0$, 0.08 eV for $S_1$, 0.12 eV for $S_2$. Noteworthy, from a direct comparison with the free-energy simulations of the 5-mer of unsubstituted fluorene and the 5-mer of 9,9-dioctylfluorene, the helix inversion is predicted to be more favored in alkyl substituted oligofluorenes where the dihedral values towards 155 degrees are more stable with respect to the unsubstituted ones.

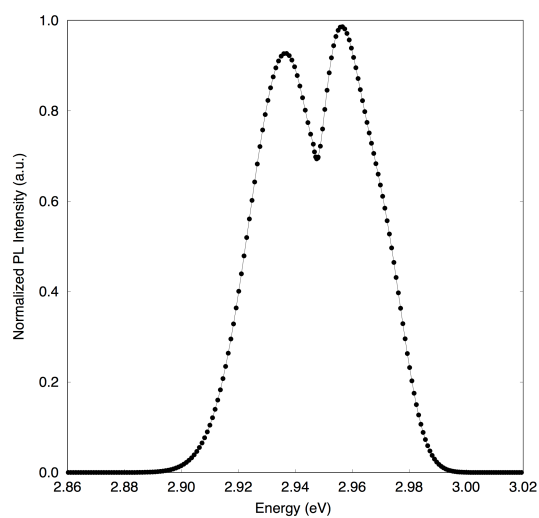

**Figure 4.** Calculated emission spectra from the $S_1$-$S_0$ emission free-energies calculated within the REM/PBMetaD/FEP scheme. The intensity has been predicted via the calculation of the transition dipole moments at the ZIndo/S level, from the coordinates belonging to the $S_1$ state sampled within the REM/PBMetaD/FEP scheme.

**IV. CONCLUSIONS**

In summary, we have introduced a simulation scheme that combines Hamiltonian Replica Exchange, Parallel Bias metadynamics and the free-energy perturbation theory to enhance sampling of the conformational space of a switchable 5-mer of 9,9-dioctylfluorene and at the same time to estimate the free-energy gaps between ground and excited states. The simulation framework presented here is particularly indicated when thermal effects accompany the irradiation process. Two different photoswitchable pathways were discussed. Specifically, with an excitation to the first excited state, a dihedral switching from the absolute free-energy minimum identifying a conformation with dihedrals at 155 degrees was predicted. From the $S_1$ state, a further transition can drive the 5-mer of 9,9-dioctylfluorene away from the global free-energy minimum, reaching the $S_2$ state and allowing the oligomer to decay in $\pm 44$ degrees twisted conformations.

Furthermore, the method predicted with good accuracy the emission spectrum and pointed out the highest stability in the ground state of a $5_2$ helix conformation compared to a $5_1$ helix one.

This biasing scheme presented here is general and can be used to predict large-scale rearrangements occurring through light-matter interactions, particularly when thermal effects are relevant for the occurrence of photoswitchable twisted-coplanar transitions in excited states.

**SUPPLEMENTARY MATERIAL**

See the supplementary material for the evolution of all dihedral angles during the time (Figure S1-S3), replica diffusions during the simulations (Figure S4), the free-energy profiles as a function of $\varphi_3, \varphi_4$ (Figure S5), the benchmarked potential energy scans (Figure S6-S8), the values of potential energy scans (Tables S1,S2) and ESP charges (Tables S3 and S4).

**DATA AVAILABILITY STATEMENT**

All the GROMACS input and topology files as well as the PLUMED input files used in this study are available on PLUMED-NEST (www.plumed-nest.org), under accession id plumID:21.008.

**ACKNOWLEDGMENTS**

The ISCRA supercomputing initiative is acknowledged for computational time provided.
This research was funded by Italian Ministry of Research under the program PRIN2017 no. 2017WBZFHL_003 (AP).